\documentclass[11pt]{article}
\usepackage[margin=1in]{geometry}

\usepackage{amsmath,amssymb}
\usepackage{graphicx}
\usepackage{tikz-cd}
\usepackage{float}
\usepackage{multicol}
\usepackage{url}
\usetikzlibrary{arrows.meta,calc,fit,positioning}
\definecolor{energyblue}{RGB}{46,98,168}
\definecolor{energyfill}{RGB}{236,244,253}
\definecolor{varioteal}{RGB}{37,136,122}
\definecolor{variofill}{RGB}{235,247,243}
\definecolor{crpspurple}{RGB}{117,78,168}
\definecolor{crpsfill}{RGB}{244,240,250}
\definecolor{targetorange}{RGB}{221,141,62}
\definecolor{targetfill}{RGB}{252,236,214}
\definecolor{weightgray}{RGB}{120,126,138}
\tikzset{
score source/.style={circle, draw=black!40, line width=0.9pt, minimum size=7.5mm, inner sep=0pt, fill=white},
score target/.style={score source, draw=targetorange!85!black, line width=1.1pt, fill=targetfill},
score energy edge/.style={-{Latex[length=2.1mm,width=1.8mm]}, line width=1.1pt, draw=energyblue!80},
score variogram edge/.style={-{Latex[length=2.1mm,width=1.8mm]}, line width=1.1pt, draw=varioteal!85},
score crps edge/.style={-{Latex[length=2.1mm,width=1.8mm]}, line width=1.1pt, draw=crpspurple!85},
score panel/.style={rounded corners=7pt, line width=0.8pt, draw=black!8},
score subtitle/.style={font=\footnotesize, text=black!60},
score energy value/.style={rounded corners=3pt, fill=energyblue!10, text=energyblue!85!black, font=\scriptsize, inner xsep=4pt, inner ysep=2pt},
score variogram value/.style={rounded corners=3pt, fill=varioteal!10, text=varioteal!85!black, font=\scriptsize, inner xsep=4pt, inner ysep=2pt},
score crps value/.style={rounded corners=3pt, fill=crpspurple!10, text=crpspurple!85!black, font=\scriptsize, inner xsep=4pt, inner ysep=2pt},
score footer/.style={rounded corners=4pt, fill=white, draw=black!6, align=left, font=\footnotesize, inner xsep=6pt, inner ysep=5pt, text width=7.0cm},
score chip/.style={rounded corners=4pt, font=\scriptsize\bfseries, inner xsep=5pt, inner ysep=3pt},
score weight/.style={font=\scriptsize, text=weightgray},
}

\newcommand{\includeSkillFigure}[2][\textwidth]{\includegraphics[width=#1,trim=0 0 0 28bp,clip]{#2}}
\newcommand{\includeSpectraFigure}[1]{\includegraphics[width=\textwidth,trim=0 0 0 42bp,clip]{#1}}

\title{On the sensitivity of machine-learned probabilistic weather forecast models to scale-aware scoring rules}
\author{Simon Lang, Martin Leutbecher, Sam Hatfield}
\date{\today}

\begin{document}
\maketitle

\begin{abstract}
Probabilistic forecast models can be machine-learned from data using loss functions based on scoring rules such as the Continuous Ranked Probability Score (CRPS). This note summarises a preliminary study comparing versions of AIFS-CRPS, a global weather forecast model, trained with different univariate and multivariate scoring rules that aim to explicitly represent scale-awareness in the loss function. In the first part, we compare the (almost) fair CRPS, a fair global energy score, and a graph energy score based on node neighbourhoods. Across standard verification metrics, forecast skill is broadly similar. In the extratropics we find only small differences, while in the tropics the graph energy score setup performs somewhat better and the global energy score shows some degradation. These results suggest that multivariate scores are a viable alternative to CRPS-based training for global machine-learned weather forecasting. In the second part of the study, we analyse how different scoring rules and scale-aware loss constraints shape the spectra of forecast fields. It is apparent that any form of explicit scale-awareness improves realism. Here, the largest differences are likely associated with different effective weights per scale. All scores are implemented in the Anemoi framework.
\end{abstract}

\section{Introduction}
Machine-learned weather forecasting has advanced rapidly in recent years, with probabilistic global models now reaching competitive skill at a fraction of the cost of traditional ensemble prediction systems. A common approach for training ensemble models is to optimize a proper scoring rule as the training loss, for example, the continuous ranked probability score (CRPS) and its fair or almost fair variants \cite{ferro2014fair,leutbecher2019ensemble}. AIFS-CRPS \cite{lang2024aifscrpsensembleforecastingusing} proposes end-to-end CRPS-based training for global fully machine-learned probabilistic weather models, an approach that has since also been used in recent global models such as FGN \cite{alet2025skillful}, FourCastNet~3 \cite{bonev2025fourcastnet3}, Huracan \cite{ni2025huracan}, and others \cite{cachay2026ucastsurprisinglysimpleefficient, diaconu2026otterweatherskillfulcomputationally, perkins2026hiroacefastskillfulai}. The approach is also being applied for regional and high-resolution forecasting, including limited-area \cite{larsson2025crpslam} and stretched-grid modelling \cite{nordhagen2025highresolution}. Meanwhile, proper-score training has now also been explored outside machine-learned weather forecasting, for example in climate downscaling \cite{schillinger2025enscale}, for training probabilistic neural operators \cite{buelte2025probabilisticneuraloperators} and learned probabilistic filtering for data assimilation \cite{bach2026learningprobabilisticfiltersstrictly}. Graph neural networks and multivariate score-based objectives have also been used for ensemble post-processing \cite{lakatos2026composite}.

Here, we address the question whether the recent success of state of the art probabilistic machine-learned weather forecasting is tied specifically to pointwise scores such as CRPS, or whether similar results can also be obtained with multivariate scoring rule optimisation, similar to \cite{pacchiardi2024probabilistic}. We compare the forecast skill of models trained with different loss objectives, with the aim to test whether multivariate spatial training objectives can match the large-scale forecast skill obtained with CRPS-based training. Furthermore, we assess the impact of  different approaches that inject scale awareness \cite{lang2025multiscalelossformulationlearning} into the loss objective on  spectra of forecast fields. Because spectra characterize how variance is distributed across spatial scales, they provide a useful diagnostic of whether forecast fields reproduce the scale-dependent structure of the atmosphere.

\section{Probabilistic Scores}\label{probscores}
We consider an ensemble of size $M$, with ensemble members $x^{(1)}, \dots, x^{(M)}$ and verifying target $y$.
Fair variants correct the finite-ensemble bias of empirical score estimates that arises from using only $M$ ensemble members \cite{ferro2014fair}.
We use $m$ and $\ell$ for ensemble-member indices, $g$ for generic grid-point indices, $n$ for a destination graph node, $j$ for a source graph node, $i$ for scale indices, $\kappa$ for spectral-mode indices, and $\mathcal{G}$ for the neighbourhood graph.
For scalar scores, $x^{(m)}$ and $y$ are single values at one grid point.
For multivariate scores of fields, $x^{(m)}$ and $y$  are vectors containing values at the grid points of the field. All scores are defined for a single forecast step and a single output variable. Hence, in the following, the energy score, graph energy score, graph variogram score, and graph edge energy score are multivariate over space or local edge differences for that variable, not across different output variables.
When a score is aggregated over the discrete forecast grid, we use spatial quadrature weights $\omega_g>0$, normalized so that $\sum_{g=1}^{G}\omega_g=1$.

\section*{CRPS}
For scalar quantities, the standard (unfair) CRPS can be written in kernel form \cite{gneiting2007proper} as
\[
\mathrm{CRPS}(x^{(1:M)}, y)
= \frac{1}{M} \sum_{m=1}^{M} \left|x^{(m)} - y\right|
- \frac{1}{2M^2}\sum_{m=1}^{M}\sum_{\ell=1}^{M} \left|x^{(m)} - x^{(\ell)}\right|.
\]
The fair version is then \cite{ferro2014fair,leutbecher2019ensemble}:
\[
\mathrm{fCRPS}(x^{(1:M)}, y)
= \frac{1}{M} \sum_{m=1}^{M} \left|x^{(m)} - y\right|
- \frac{1}{2M(M-1)}\sum_{m=1}^{M}\sum_{\substack{\ell=1\\ \ell\ne m}}^{M} \left|x^{(m)} - x^{(\ell)}\right|.
\]
However, the fCRPS has a degeneracy: When all but one ensemble member are equal to the observation, the remaining member is unconstrained by the score.
Following \cite{lang2024aifscrpsensembleforecastingusing}, an almost fair variant can be defined as a convex combination of the standard and fair scores,
\[
\mathrm{afCRPS}_{\alpha}(x^{(1:M)}, y)
= \alpha \, \mathrm{fCRPS}(x^{(1:M)}, y)
+ (1-\alpha)\,\mathrm{CRPS}(x^{(1:M)}, y),
\]
with $\alpha \in [0,1]$.
Using the kernel representation, this becomes
\[
\mathrm{afCRPS}_{\alpha}(x^{(1:M)}, y)
= \frac{1}{M} \sum_{m=1}^{M} \left|x^{(m)} - y\right|
- \frac{1}{2}\frac{1-\epsilon}{M(M-1)}
\sum_{m=1}^{M}\sum_{\substack{\ell=1\\ \ell\ne m}}^{M} \left|x^{(m)} - x^{(\ell)}\right|,
\qquad
\epsilon = \frac{1-\alpha}{M}.
\]
Thus $\alpha=1$ recovers the fair CRPS, while $\alpha=0$ recovers the standard CRPS.
For a field, the pointwise score is aggregated with the spatial weights,
\[
\mathcal{L}_{\mathrm{afCRPS},\alpha}
=
\sum_{g=1}^{G}
\omega_g\,
\mathrm{afCRPS}_{\alpha}
\left(x_g^{(1:M)}, y_g\right).
\]

\section*{Energy Score}
The energy score is a multivariate score \cite{gneiting2007proper}. Here, we compute it over the whole spatial field of a single output variable.
For a field difference $d \in \mathbb{R}^{G}$ over $G$ grid points, it uses the spatially weighted Euclidean norm
\[
\|d\|_{W}
=
\left( \sum_{g=1}^{G} \omega_g \, d_g^2 \right)^{1/2},
\]
where $\omega_g$ are the spatial grid weights defined above.
The standard (unfair) energy score is
\[
\mathrm{ES}(x^{(1:M)}, y)
= \frac{1}{M} \sum_{m=1}^{M} \left\|x^{(m)} - y\right\|_{W}
- \frac{1}{2M^2} \sum_{m=1}^{M} \sum_{\ell=1}^{M} \left\|x^{(m)} - x^{(\ell)}\right\|_{W}
\]
and the fair energy score \cite{ferro2014fair} is given by
\[
\mathrm{fES}(x^{(1:M)}, y)
= \frac{1}{M} \sum_{m=1}^{M} \left\|x^{(m)} - y\right\|_{W}
- \frac{1}{2M(M-1)} \sum_{m=1}^{M} \sum_{\substack{\ell=1\\ \ell\ne m}}^{M} \left\|x^{(m)} - x^{(\ell)}\right\|_{W}.
\]
Compared with the CRPS, this score treats the whole field as one vector, so it measures whether the ensemble reproduces the joint spatial state and not only the marginal distribution at each grid point.

\section*{Graph Energy Score}
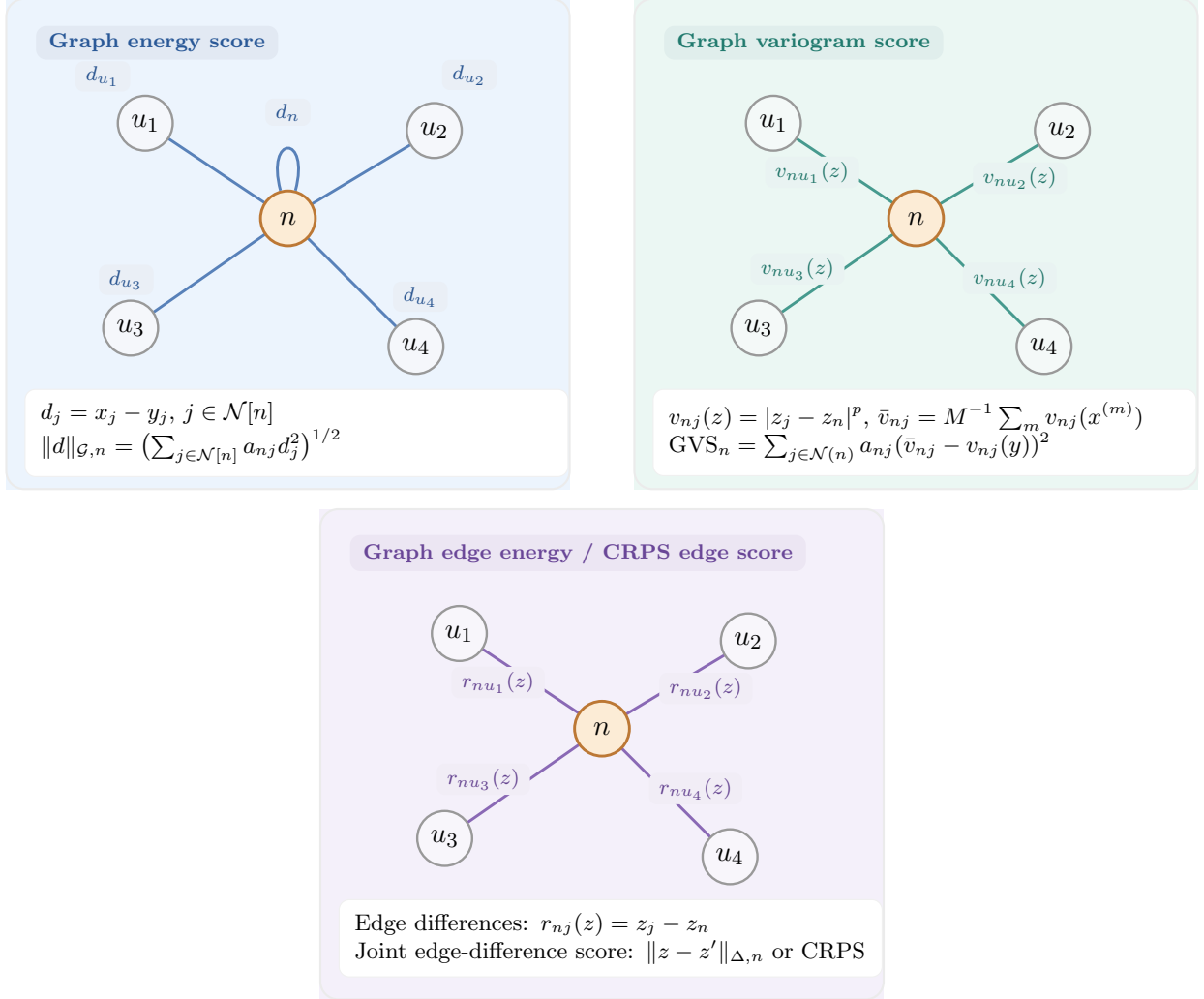
\begin{figure}[H]
\centering
\begin{minipage}[t]{0.48\textwidth}
\centering
\begin{tikzpicture}[x=1cm,y=1cm]
    \fill[energyfill] (-3.85,-3.55) rectangle (3.85,3.15);
    \draw[score panel] (-3.85,-3.55) rectangle (3.85,3.15);
    \node[score chip, fill=energyblue!14, text=energyblue!90!black, anchor=west] at (-3.45,2.55) {Graph energy score};
    \coordinate (nE) at (0.0,0.15);
    \coordinate (j1E) at (-1.95,1.45);
    \coordinate (j2E) at (2.0,1.35);
    \coordinate (j3E) at (-2.15,-1.35);
    \coordinate (j4E) at (1.75,-1.6);

    \foreach \src in {j1E,j2E,j3E,j4E} {
        \draw[score energy edge] (\src) -- (nE);
    }

    \node[score source, fill=energyblue!5] at (j1E) {$u_1$};
    \node[score source, fill=energyblue!5] at (j2E) {$u_2$};
    \node[score source, fill=energyblue!5] at (j3E) {$u_3$};
    \node[score source, fill=energyblue!5] at (j4E) {$u_4$};
    \node[score target] (targetE) at (nE) {$n$};
    \path[draw=energyblue!80, line width=1.1pt] (targetE) edge[loop above, min distance=8mm, looseness=5] (targetE);

    \node[score energy value] at ($(j1E)+(-0.58,0.64)$) {$d_{u_1}$};
    \node[score energy value] at ($(j2E)+(0.48,0.76)$) {$d_{u_2}$};
    \node[score energy value] at ($(j3E)+(-0.06,0.66)$) {$d_{u_3}$};
    \node[score energy value] at ($(j4E)+(0.06,0.68)$) {$d_{u_4}$};
    \node[score energy value] at ($(nE)+(0.00,1.45)$) {$d_n$};

    \node[score footer, anchor=north west] at (-3.60,-2.18) {$d_j=x_j-y_j$, $j\in\mathcal{N}[n]$\\$\|d\|_{\mathcal{G},n} = \bigl(\sum_{j \in \mathcal{N}[n]} a_{nj} d_j^2\bigr)^{1/2}$};
\end{tikzpicture}
\end{minipage}\hfill
\begin{minipage}[t]{0.48\textwidth}
\centering
\begin{tikzpicture}[x=1cm,y=1cm]
    \fill[variofill] (-3.85,-3.55) rectangle (3.85,3.15);
    \draw[score panel] (-3.85,-3.55) rectangle (3.85,3.15);
    \node[score chip, fill=varioteal!14, text=varioteal!90!black, anchor=west] at (-3.45,2.55) {Graph variogram score};
    \coordinate (nV) at (0.0,0.15);
    \coordinate (j1V) at (-1.95,1.45);
    \coordinate (j2V) at (2.0,1.35);
    \coordinate (j3V) at (-2.15,-1.35);
    \coordinate (j4V) at (1.75,-1.6);

    \foreach \src in {j1V,j2V,j3V,j4V} {
        \draw[score variogram edge] (\src) -- (nV);
    }

    \node[score source, fill=varioteal!5] at (j1V) {$u_1$};
    \node[score source, fill=varioteal!5] at (j2V) {$u_2$};
    \node[score source, fill=varioteal!5] at (j3V) {$u_3$};
    \node[score source, fill=varioteal!5] at (j4V) {$u_4$};
    \node[score target] at (nV) {$n$};

    \node[score variogram value] at (-1.42,0.78) {$v_{n u_1}(z)$};
    \node[score variogram value] at (1.42,0.70) {$v_{n u_2}(z)$};
    \node[score variogram value] at (-1.62,-0.54) {$v_{n u_3}(z)$};
    \node[score variogram value] at (1.28,-0.67) {$v_{n u_4}(z)$};

    \node[score footer, anchor=north west] at (-3.60,-2.18) {$v_{nj}(z)=|z_j-z_n|^p$, $\bar v_{nj}=M^{-1}\sum_m v_{nj}(x^{(m)})$\\$\mathrm{GVS}_n=\sum_{j\in\mathcal{N}(n)} a_{nj}(\bar v_{nj}-v_{nj}(y))^2$};
\end{tikzpicture}
\end{minipage}

\medskip

\begin{minipage}[t]{0.48\textwidth}
\centering
\begin{tikzpicture}[x=1cm,y=1cm]
    \fill[crpsfill] (-3.85,-3.55) rectangle (3.85,3.15);
    \draw[score panel] (-3.85,-3.55) rectangle (3.85,3.15);
    \node[score chip, fill=crpspurple!14, text=crpspurple!90!black, anchor=west] at (-3.45,2.55) {Graph edge energy / CRPS edge score};
    \coordinate (nEC) at (0.0,0.15);
    \coordinate (j1EC) at (-1.95,1.45);
    \coordinate (j2EC) at (2.0,1.35);
    \coordinate (j3EC) at (-2.15,-1.35);
    \coordinate (j4EC) at (1.75,-1.6);

    \foreach \src in {j1EC,j2EC,j3EC,j4EC} {
        \draw[score crps edge] (\src) -- (nEC);
    }

    \node[score source, fill=crpspurple!5] at (j1EC) {$u_1$};
    \node[score source, fill=crpspurple!5] at (j2EC) {$u_2$};
    \node[score source, fill=crpspurple!5] at (j3EC) {$u_3$};
    \node[score source, fill=crpspurple!5] at (j4EC) {$u_4$};
    \node[score target] at (nEC) {$n$};

    \node[score crps value] at (-1.42,0.78) {$r_{n u_1}(z)$};
    \node[score crps value] at (1.42,0.70) {$r_{n u_2}(z)$};
    \node[score crps value] at (-1.62,-0.54) {$r_{n u_3}(z)$};
    \node[score crps value] at (1.28,-0.67) {$r_{n u_4}(z)$};

    \node[score footer, anchor=north west] at (-3.60,-2.18) {Edge differences: $r_{nj}(z)=z_j-z_n$\\Joint edge-difference score: $\|z-z'\|_{\Delta,n}$ or CRPS};
\end{tikzpicture}
\end{minipage}
\caption{Different scores for a single destination node $n$: the graph energy score (top left) evaluates forecast errors on the closed neighbourhood $\mathcal{N}[n]$, which includes the destination node itself, and aggregates them through a weighted neighbourhood norm. The graph variogram score (top right) evaluates squared differences between forecast and observed variogram values on the open neighbourhood $\mathcal{N}(n)$. The edge-scores (bottom) use edge differences, either as one local vector for the graph edge energy score or as scalar quantities for the CRPS edge score. Final scores are then obtained by spatial aggregation over destination nodes with weights $\omega_n$. In the schematic, $u_1,\dots,u_4$ denote example source nodes in $\mathcal{N}(n)$.}
\label{fig:neighbourhood-score-constructions}
\end{figure}

The graph energy score localizes the energy score by replacing the norm computed over the full spatial field with a weighted neighbourhood norm defined on a graph.
The motivation is similar to the patched energy score of \cite{pacchiardi2024probabilistic}, where a multivariate score is evaluated on localized subsets and then aggregated. This kind of localization is intended to make the score more sensitive to local multivariate relationships than a purely global energy score. In the weather forecasting experiments of \cite{pacchiardi2024probabilistic}, patched energy scores performed best among the multivariate scoring rules they considered, which motivates exploring a graph-based localization in the present setting. Compared with earlier patch-based localization approaches, we use local graph neighbourhoods on the data grid, which provide a more flexible way to localize multivariate scoring rules. An advantage of the graph-based approach is that it does not rely on fixed rectangular patches on a regular grid. Instead, it only requires a neighbourhood graph and associated weights, so the same score can in principle be applied on irregular grids, sparse spatial meshes, or sparse observation networks.
For a target node $n$, let $\mathcal{N}(n)$ be its open incoming neighbourhood, i.e. the source nodes connected to $n$ excluding $n$ itself, and let
\[
\mathcal{N}[n] = \mathcal{N}(n) \cup \{n\}
\]
be the corresponding closed neighbourhood.
For the graph energy score we use the closed neighbourhood $\mathcal{N}[n]$, with edge weights $a_{nj}$ for $j \in \mathcal{N}[n]$.
The self-node contribution is included. 
We use normalised uniform  edge weights $a_{nj}=a_n=1/{\vert \mathcal{N}[n]\vert}$, resulting in a unit sum
\[
\sum_{j \in \mathcal{N}[n]} a_{nj} = 1.
\]
At destination node $n$, the weighted neighbourhood norm is
\[
\|d\|_{\mathcal{G},n} = \left( \sum_{j \in \mathcal{N}[n]} a_{nj} \, d_j^2 \right)^{1/2}.
\]
The fair graph energy score at target node $n$ is then
\[
\mathrm{fGES}_n(x^{(1:M)}, y)
= \frac{1}{M} \sum_{m=1}^{M} \left\|x^{(m)} - y\right\|_{\mathcal{G},n}
- \frac{1}{2M(M-1)} \sum_{m=1}^{M} \sum_{\substack{\ell=1\\ \ell\ne m}}^{M} \left\|x^{(m)} - x^{(\ell)}\right\|_{\mathcal{G},n},
\]
and the corresponding spatially aggregated graph energy score is
\[
\mathrm{fGES}_{\mathrm{graph}}(x^{(1:M)}, y)
=
\sum_{n=1}^{G}\omega_n\,
\mathrm{fGES}_n(x^{(1:M)}, y).
\]
Instead of scoring the full domain at once, the graph energy score scores each node using only the structure in its prescribed neighbourhood, which makes the loss more sensitive to local spatial organization. Across the full domain, we evaluate this local score at every destination node $n$. This is equivalent to a sliding-window score evaluated at every node, with maximal overlap between adjacent windows. Here, however, each local window is defined by the closed graph neighbourhood $\mathcal{N}[n]$ and weights $a_{nj}$ rather than by a fixed rectangular stencil. Figure~\ref{fig:neighbourhood-score-constructions} (top-left panel) shows the construction at one destination node.

The energy score is strictly proper. The graph energy score, however, may fail to be strictly proper because different distributions can lead to the same local score even though they differ in their long-range dependence. This can be illustrated by simple counterexamples. Strict propriety may be recovered under additional assumptions, but it is not guaranteed in general. We therefore follow \cite{pacchiardi2024probabilistic} and combine the proper graph energy score with a weak global anchor in the form of the fair energy score. This kind of combination of proper scoring rules preserves propriety for non-negative weights and can be used to target complementary features of multivariate forecasts \cite{pic2025proper}. The global component captures dependencies that are invisible to the local score and makes the combined score strictly proper.

\section*{Graph Variogram Score}
The graph variogram score is a variant of the variogram score \cite{scheuerer2015variogram}. It uses the same fully overlapping node neighbourhood formulation as the graph energy score, but replaces node values by per-edge variogram differences. For an edge from source node $j$ to destination node $n$, the variogram transform is
\[
v_{nj}(z) = \left| z_j - z_n \right|^p,
\]
where $p>0$ is the variogram exponent.
For the graph variogram score, we use the open neighbourhood $\mathcal{N}(n)$, so the self-edge $j=n$ is not included. The edge weights are uniform and normalised to unit sum $a_{nj}=1/{\vert \mathcal{N}(n)\vert}$, i.e.\  $\sum_{j \in \mathcal{N}(n)} a_{nj} = 1$. Figure~\ref{fig:neighbourhood-score-constructions} (top-right panel) shows the corresponding formulation on the same neighbourhood graph.

For a single node $n$, the standard (unfair) graph variogram score aggregates squared differences between the observed variogram and the ensemble-mean variogram:
\[
\mathrm{GVS}_n(x^{(1:M)}, y)
= \sum_{j \in \mathcal{N}(n)} a_{nj}
\left(
\frac{1}{M}\sum_{m=1}^{M} v_{nj}(x^{(m)}) - v_{nj}(y)
\right)^2.
\]
To derive the fair score, we define the weighted squared distance between two local variogram vectors as
\[
D_n^2(x,x')
=
\sum_{j\in\mathcal{N}(n)}
a_{nj}
\left(
v_{nj}(x)-v_{nj}(x')
\right)^2.
\]

Then we obtain the fair graph variogram score:
\[
\mathrm{fGVS}_n(x^{(1:M)}, y)
=
\frac{1}{M}\sum_{m=1}^M D_n^2(x^{(m)}, y)
-
\frac{1}{2M(M-1)}
\sum_{m=1}^M
\sum_{\substack{\ell=1\\ \ell\ne m}}^M
D_n^2(x^{(m)}, x^{(\ell)}).
\]

The corresponding spatially aggregated graph variogram score is
\[
\mathrm{fGVS}_{\mathrm{graph}}(x^{(1:M)}, y)
=
\sum_{n=1}^{G}\omega_n\,
\mathrm{fGVS}_n(x^{(1:M)}, y).
\]
The score measures whether the ensemble reproduces local spatial variability.

\section*{Graph Edge Energy Score}
The graph edge energy score uses the same open neighbourhood as the graph variogram score, but uses the edge differences rather than applying the variogram transform. For an edge from source node $j$ to destination node $n$, we define
\[
r_{nj}(z) = z_j - z_n,
\qquad j\in\mathcal{N}(n).
\]
Using the uniform normalized edge weights with $\sum_{j\in\mathcal{N}(n)} a_{nj}=1$, the edge-differences feature vector is
\[
r_n(z)
=
\left(
\sqrt{a_{nj}}\,r_{nj}(z)
\right)_{j\in\mathcal{N}(n)}.
\]
The edge-difference distance is defined as
\[
\|z-z'\|_{\Delta,n}
=
\|r_n(z)-r_n(z')\|
=
\left(
\sum_{j\in\mathcal{N}(n)}
a_{nj}
\left[
(z_j-z_n)-(z'_j-z'_n)
\right]^2
\right)^{1/2}.
\]
The fair graph edge energy score at node $n$ applies the fair energy score to this vector of edge differences:
\[
\mathrm{fGEES}_n(x^{(1:M)}, y)
= \frac{1}{M} \sum_{m=1}^{M} \left\|x^{(m)} - y\right\|_{\Delta,n}
- \frac{1}{2M(M-1)} \sum_{m=1}^{M} \sum_{\substack{\ell=1\\ \ell\ne m}}^{M}
\left\|x^{(m)} - x^{(\ell)}\right\|_{\Delta,n}.
\]
The corresponding spatially aggregated graph edge energy score is
\[
\mathrm{fGEES}(x^{(1:M)}, y)
=
\sum_{n=1}^{G}\omega_n\,
\mathrm{fGEES}_n(x^{(1:M)}, y).
\]
The score compares the joint vector of edge differences in the local neighbourhood $\mathcal{N}(n)$. Compared with the graph variogram score, it preserves the sign of the edge differences.

\section*{CRPS Edge Score}
The CRPS edge score uses the same open neighbourhood $\mathcal{N}(n)$, normalized weights $a_{nj}$, and edge differences $r_{nj}(z)$ as the graph edge energy score, but scores each edge difference as a scalar quantity. For the almost fair edge CRPS, the scalar score on edge $(j,n)$ is
\[
\begin{aligned}
\mathrm{ECRPS}_{\alpha,nj}(x^{(1:M)}, y)
&=
\frac{1}{M} \sum_{m=1}^{M}
\left|r_{nj}(x^{(m)}) - r_{nj}(y)\right|
\\
&\quad
- \frac{1}{2}\frac{1-\epsilon}{M(M-1)}
\sum_{m=1}^{M}\sum_{\substack{\ell=1\\ \ell\ne m}}^{M}
\left|r_{nj}(x^{(m)}) - r_{nj}(x^{(\ell)})\right|,
\qquad
\epsilon = \frac{1-\alpha}{M}.
\end{aligned}
\]
At each destination node $n$, the score is then aggregated over the source nodes $j$
\[
\mathrm{ECRPS}_{\alpha,n}(x^{(1:M)}, y)
=
\sum_{j\in\mathcal{N}(n)} a_{nj}\,
\mathrm{ECRPS}_{\alpha,nj}(x^{(1:M)}, y).
\]
The spatially aggregated edge CRPS is then given by
\[
\mathrm{ECRPS}_{\alpha}(x^{(1:M)}, y)
=
\sum_{n=1}^{G}\omega_n\,
\mathrm{ECRPS}_{\alpha,n}(x^{(1:M)}, y).
\]
The edge CRPS assesses whether, for each edge, the marginal ensemble distribution of edge differences matches the observed difference.
Figure~\ref{fig:neighbourhood-score-constructions} (bottom panel) shows how the edge-difference are defined for the graph edge energy score and the edge CRPS.
Because both edge scores, like the graph variogram score, depend only on differences along graph edges, they should be combined with a score that
constrains the absolute node values.

\section*{Multi-Scale Score definition}
We use the multi-scale loss formulation of \cite{lang2025multiscalelossformulationlearning}. For scalar fields on a spatial manifold $\mathcal{M}$, let
\[
x^{(m)}:\mathcal{M}\to\mathbb{R},
\qquad
y:\mathcal{M}\to\mathbb{R},
\]
denote the $m$-th ensemble member and target field. In the discrete case, the integrals below are replaced by the corresponding weighted sums over grid points.

For a pointwise scalar scoring rule $\mathcal{S}$, the scale-unaware loss is
\[
\mathcal{L}_{\text{scale-unaware}}
=
c\int_{\mathcal{M}}
\mathcal{S}\!\left(
\left[x^{(m)}(q)\mid m=1,\dots,M\right],
y(q)
\right)\,\mathrm{d}\mu(q),
\]
where $\mu$ is a measure on $\mathcal{M}$ and $c$ is a normalization constant. This applies the score at each location $q\in\mathcal{M}$, followed by spatial aggregation. It is not scale-aware because the score only sees the marginal ensemble distribution at each location.

The multi-scale loss introduces ordered smoothing operators $D_1,\dots,D_{K-1}$, where $D_i$ smooths more strongly than $D_{i+1}$. These operators partition a field into $K$ residual scale bands:
\[
z_{\mathrm{scale}\,1}=D_1 z,
\]
\[
z_{\mathrm{scale}\,i}=D_i z-D_{i-1}z,
\qquad i=2,\dots,K-1,
\]
\[
z_{\mathrm{scale}\,K}=z-D_{K-1}z.
\]
The first band contains the coarsest component, while later bands contain progressively finer residuals. This approach is similar to a Laplacian-pyramid or Laplacian-cascade decomposition \cite{burt1983laplacian}: successive low-pass filtered fields define residual bands that are localized in scale. Following \cite{lang2025multiscalelossformulationlearning}, we use the same idea as a scale decomposition for scoring. The smoothing operators may be implemented as spatial kernel smoothers with decreasing width, or as spectral filters when a suitable transform is available.

The $K$-scale loss is then the weighted sum of the loss on each scale band:
\[
\mathcal{L}_{K\text{-scale}}
=
\sum_{i=1}^{K}
\zeta_i\,c
\int_{\mathcal{M}}
\mathcal{S}\!\left(
\left[x^{(m)}_{\mathrm{scale}\,i}(q)\mid m=1,\dots,M\right],
y_{\mathrm{scale}\,i}(q)
\right)\,\mathrm{d}\mu(q),
\]
with scale weight $\zeta_i>0$. The approach is score-agnostic: once the fields are decomposed into scale bands, any suitable scoring rule can be applied separately to each band.

\section*{Scores in spectral space}
Another approach to introduce scale awareness is via the application of scoring rules in spectral space. This proceeds by 
transforming each scalar forecast field to a spectral representation. Let $\mathcal{T}$ denote a spectral transform and write
\[
\widehat{x}^{(m)}_\kappa = (\mathcal{T}x^{(m)})_\kappa,
\qquad
\widehat{y}_\kappa = (\mathcal{T}y)_\kappa,
\]
where a hat denotes the (complex) coefficient, or mode, of the spectral representation of the corresponding variable at wavenumber $\kappa$. The score is evaluated mode by mode and then aggregated over modes.

The exact kind of transform $\mathcal{T}$ used depends on the nature of the grid on which the score is being calculated. For a global grid with Gaussian latitudes, a transform based on spherical harmonics is an obvious choice, for example. This, and other spectral transforms which operate on two-dimensional fields, gives a spectral decomposition with two wavenumber indices (total and zonal wavenumbers for a spherical harmonic decomposition). Here we do not distinguish between the two wavenumber indices and simply iterate over all wavenumber combinations with a single index, $\kappa$.

We test two variants. The spectral energy score  treats each complex coefficient as a two-dimensional real vector and scores the 2-dim vectors with the energy score. This score is sensitive to both amplitude and phase. The spectral magnitude CRPS applies the CRPS to the modulus of the spectral coefficients. The spectral magnitude CRPS constrains the distribution of spectral amplitudes, but does not penalize phase errors.

For the spectral energy score, we define
\[
z_\kappa^{(m)}
=
\left(
\mathrm{Re}\,\widehat{x}^{(m)}_\kappa,
\mathrm{Im}\,\widehat{x}^{(m)}_\kappa
\right),
\qquad
z_\kappa^y
=
\left(
\mathrm{Re}\,\widehat{y}_\kappa,
\mathrm{Im}\,\widehat{y}_\kappa
\right).
\]
The score at mode $\kappa$ is then an energy score  applied to the 2-dim vectors $z_\kappa^{(1:M)}$ and $z_\kappa^y$.
For the spectral magnitude CRPS, the scalar quantities $|\widehat{x}^{(m)}_\kappa|$ and $|\widehat{y}_\kappa|$ are scored with the CRPS.
In both cases, we are left with a vector of scores, one per spectral mode. The full spectral score is obtained by aggregating over all spectral modes $\kappa$, optionally with spectral-band weights.

\section{Experiments}
We conduct two sets of experiments. The first assesses the sensitivity of forecast skill to the use of different proper scoring rules as training objectives. The second assesses how different ways of introducing scale awareness into the loss objective affect the global spectra of forecast fields. In both sets of experiments, as stated above, all losses are computed separately for each output variable before they are aggregated across variables.
The first set of experiments is described in section~\ref{exps1}. It comprises three experiments based on the CRPS, the fair global energy score, and the fair graph energy score. The second set is described in section~\ref{exps2}. It comprises twelve experiments with different loss configurations, which we compare in terms of how they shape the spectra of the forecast fields.

\subsection{Proper score forecast skill comparison}\label{exps1}
We follow AIFS-CRPS \cite{lang2024aifscrpsensembleforecastingusing} in terms of architecture and general training configuration, but restrict resolution here to an O96 $\approx 1\deg$ model. We use the Anemoi framework (\url{https://github.com/ecmwf/anemoi-core}) for experimentation. AIFS-CRPS has an encoder–processor–decoder architecture. In our experiments, the encoder maps from the data grid to the processor grid using a graph neural network, with each data-grid node connected to its four nearest processor-grid nodes. The decoder maps from the processor grid back to the data grid, with each data-grid node connected to its eight nearest processor-grid nodes. The graph connectivity for the graph-based scores is defined by a $k$-nearest-neighbour graph with $k=16$ on the O96 reduced Gaussian grid. The training schedule uses 150{,}000 iterations at rollout 1, 30{,}000 iterations at rollout 2, and then 1{,}000 iterations at each rollout step from 3 to 12. For rollout 1 and rollout 2, we use a cosine learning-rate schedule with a warmup of 1{,}000 steps followed by decay to zero. The learning rate is $10^{-3}$ for rollout 1 and $10^{-5}$ for rollout 2. For rollout steps 3 to 12, we use a fixed learning rate of $10^{-6}$. In all cases, optimization uses AdamW with weight decay 0.1. For scores computed in spectral space, we use the spherical harmonic transform capability recently added to Anemoi to transform fields. Training uses ERA5 reanalysis data \cite{hersbach2020era5} from 1979 to 2020, and inference is performed for 2022. Here we generate 8-member ensembles to compare forecast scores.
We run three experiments. The univariate baseline experiment uses the almost fair CRPS with $\alpha = 0.95$. The second uses the fair energy score. Hence, the whole field is scored jointly through a global spatial score. The third experiment uses the fair graph energy score together with a weak global fair energy score anchor,
$\mathcal{L}_{\mathrm{graph}} = \mathrm{fGES}_{\mathrm{graph}} + 0.1\,\mathrm{fES}$.

To make score computations efficient, we rely on \texttt{torch.compile} \cite{ansel2024pytorch2} to generate fused Triton kernels \cite{tillet2019triton}.

\begin{table}[H]
  \centering
  \small
  \begin{tabular}{p{0.30\textwidth}p{0.60\textwidth}}
  \hline
  Experiment & Training objective \\
  \hline
  CRPS &
  CRPS \\
  Global energy score &
  Energy score applied per variable to the full forecast field \\
  Graph energy &
  Graph energy score with a weak global energy score anchor \\
  \hline
  \end{tabular}
  \caption{Loss objectives used in the experiments described in section~\ref{exps1}. The CRPS experiment uses the almost fair CRPS with $\alpha=0.95$. The global energy and graph energy terms use the
  corresponding fair variants described in section~\ref{probscores}}
  \label{tab:large-experiments}
\end{table}

\subsection{Impact on spectra of forecast fields}\label{exps2}
The setup follows section~\ref{exps1}, except that we use a smaller model and a shorter training schedule to reduce computational cost. We use an embedding dimension of 256 and 12 processor layers. Training has only two phases. First, the model is trained as a one-step forecast model for up to 150{,}000 optimization steps. Second, the corresponding one-step model is used to initialize rollout training. In this phase, we train for 1{,}000 optimization steps at each rollout length, increasing the rollout length one step at a time up to eight forecast steps, which corresponds to 48~hours lead time.

The loss configuration of the twelve different experiments is described in table \ref{tab:small-experiments}. The single-scale CRPS experiment and the global energy score experiment are the same as the corresponding experiments described in Sec.~\ref{exps1} apart from the cheaper configuration and modified training schedule. For multi-scale experiments, we used different weighting for the small scales of geopotential and mean sea-level pressure, to account for the reduced variance associated with small scales of these fields compared to, e.g. wind. We used a similar approach for the spectral losses with per-waveband weighting. These weighting factors are ad hoc and were chosen only to be of the right order of magnitude - estimated from the data - rather than tuned. In the almost fair CRPS experiments we always set $\alpha=0.95$.

All multi-scale experiments used four graph-based Gaussian smoothing operators on the native O96 grid, with kernel widths of approximately $100$, $200$, $400$, and $800$~km. The spectral losses used a spherical harmonic transform with truncation T191. For the weighted spectral experiments, spectral modes were grouped according to their total wavenumber $\ell$, using the bands $\ell=0$--$10$, $11$--$20$, $21$--$80$, $81$--$120$, and $121$--$191$.

\begin{table}[H]
\centering
\small
\begin{tabular}{p{0.30\textwidth}p{0.60\textwidth}}
\hline
Experiment & Training objective \\
\hline
Single-scale CRPS & CRPS \\
Multi-scale CRPS & Multi-scale version of CRPS \\
Multi-scale edge CRPS & Multi-scale CRPS combined with edge CRPS \\
Global energy score & Energy score applied per variable to the full forecast field \\
Multi-scale global energy score & Multi-scale version of the global energy score \\
Multi-scale graph energy & Multi-scale graph energy score with a weak global energy score anchor \\
Multi-scale graph edge energy & Multi-scale graph energy score combined with the graph edge energy score and a weak global energy score anchor \\
Multi-scale variogram & Multi-scale CRPS combined with the graph variogram score \\
Spectral energy score& CRPS combined with the spectral energy score \\
Spectral magnitude CRPS& CRPS combined with CRPS on spectral-coefficient magnitudes \\
Spectral energy score weighted & As spectral energy, but with band and variable weighting \\
Spectral mag.\ CRPS weighted & As spectral magnitude, but with band and variable weighting \\
\hline
\end{tabular}
\caption{Loss objectives used in the experiments described in section~\ref{exps2}. CRPS-based terms use the almost fair CRPS with $\alpha=0.95$. Energy score, graph energy, graph edge energy, and graph variogram terms use the corresponding fair variants described in section~\ref{probscores}, for further details see section~\ref{exps2}.}
\label{tab:small-experiments}
\end{table}

\section{Results}
\subsection{Forecast Skill Comparison}
We first assess how sensitive forecast skill is to the choice of the proper scoring rule used as the training objective. We compare three experiments that use the almost fair CRPS, the fair global energy score, and the fair graph energy score, respectively. The architecture, training configuration, and data are otherwise kept the same across the three experiments, as described in section~\ref{exps1}. For scoring, we interpolate the fields to a resolution of $1.5\deg$ and evaluate forecast skill using the fair CRPS.

\begin{figure}[p]
\centering
\includeSkillFigure{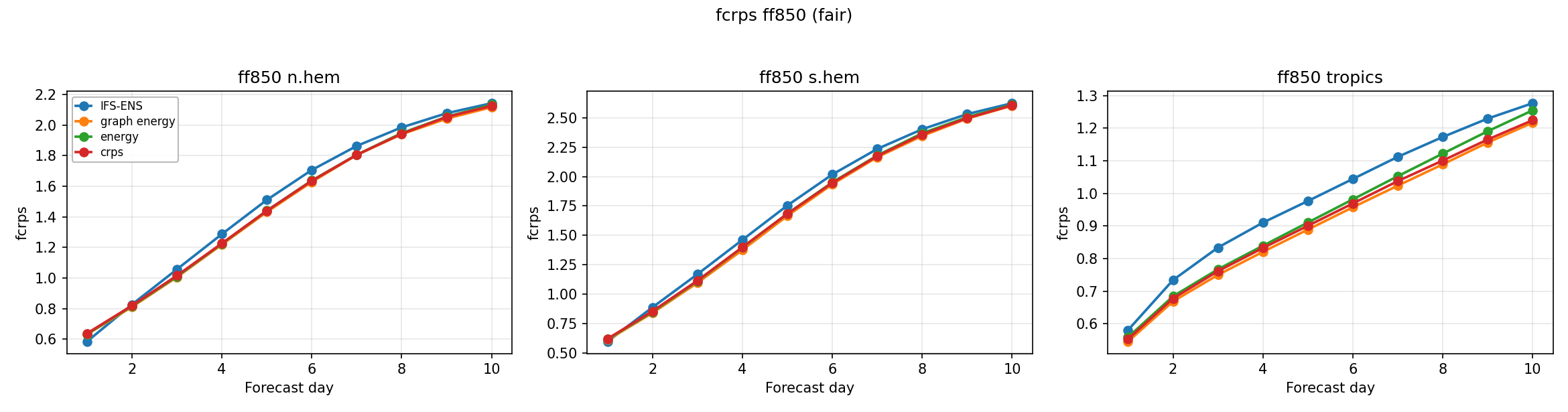}

\medskip

\includeSkillFigure{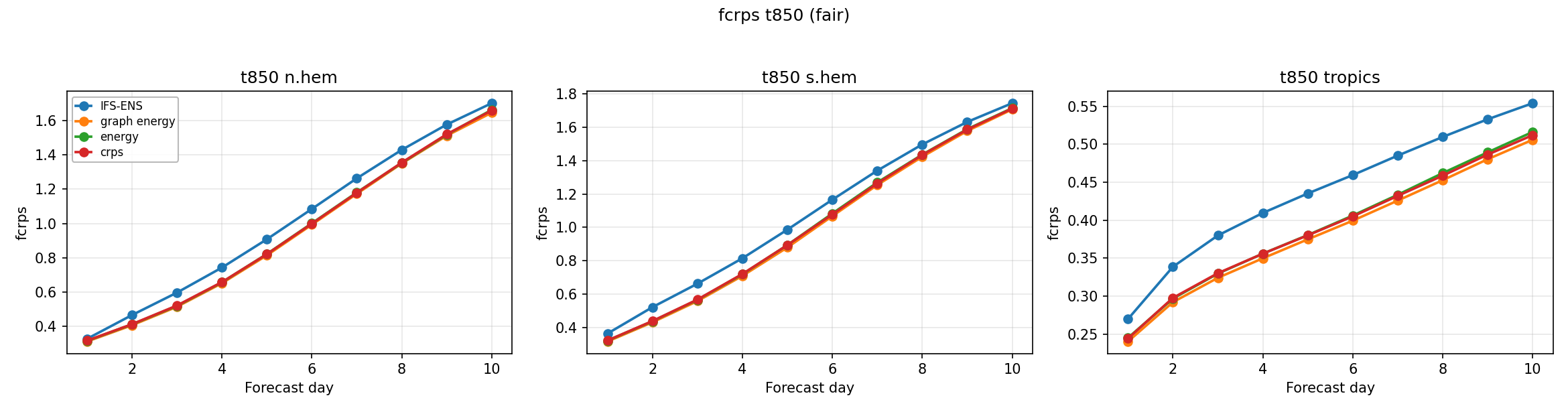}

\medskip

\includeSkillFigure[0.667\textwidth]{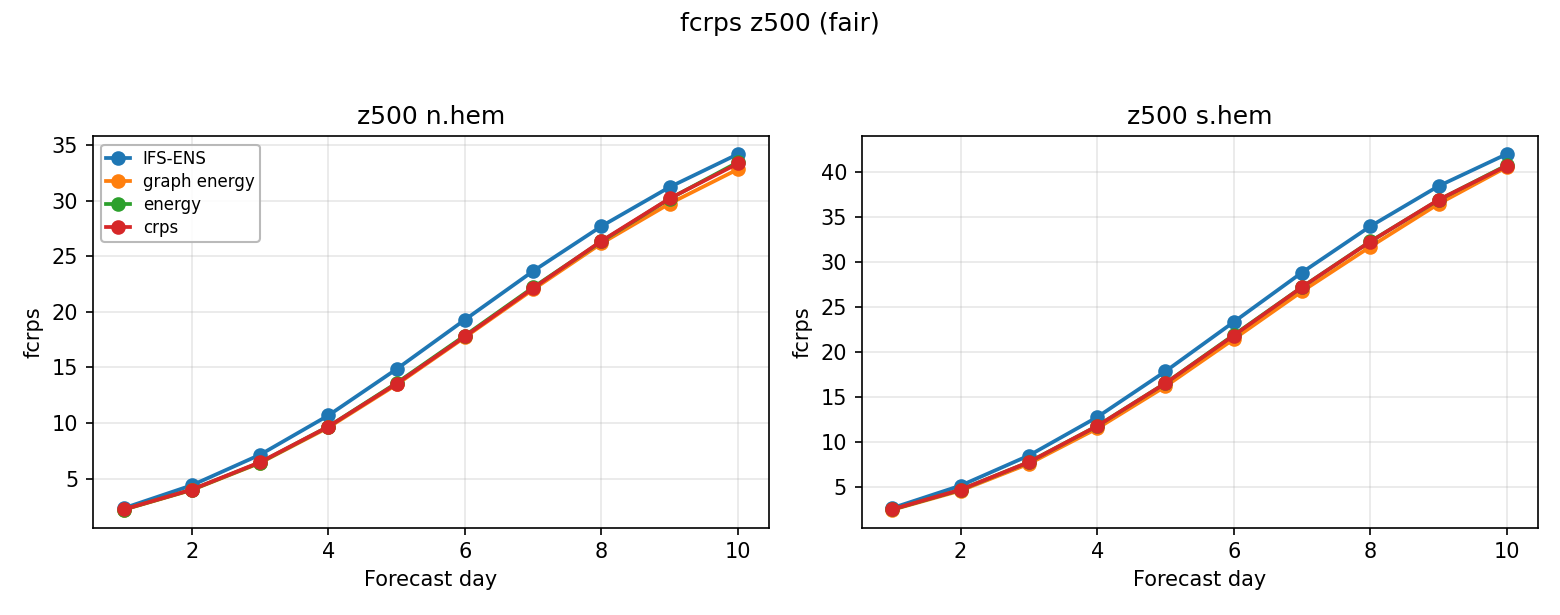}\makebox[0.333\textwidth]{}
\caption{Comparison of the CRPS, energy score, and graph energy score experiments. The top row shows wind speed at 850~hPa, the middle row temperature at 850~hPa, and the bottom row geopotential at 500~hPa. Within each row, the left panel corresponds to the Northern Hemisphere, the middle panel to the Southern Hemisphere, and the right panel to the tropics.}
\label{fig:proper-scores-all}
\end{figure}

Forecast skill is broadly similar across the experiments (see figure~\ref{fig:proper-scores-all}). In the extratropics, we do not find a noticeable difference between the three training objectives. In the tropics, however, the graph energy score experiment appears to perform best, while the energy score experiment shows some degradation relative to the other two experiments.
The conclusions are also consistent when the fair energy score or fair graph energy score is used for verification instead of the fair CRPS.

\subsection{Spectra of Forecast Fields}
We next assess how sensitive the spectra of forecast fields are to the choice of training objective. We compare the twelve experiments described in section~\ref{exps2}. Their loss objectives are based on different combinations of the probabilistic scores introduced in section~\ref{probscores}. We test how the different loss objectives affect the representation of variability at different spatial scales.

\begin{figure}[p]
\centering
\includeSpectraFigure{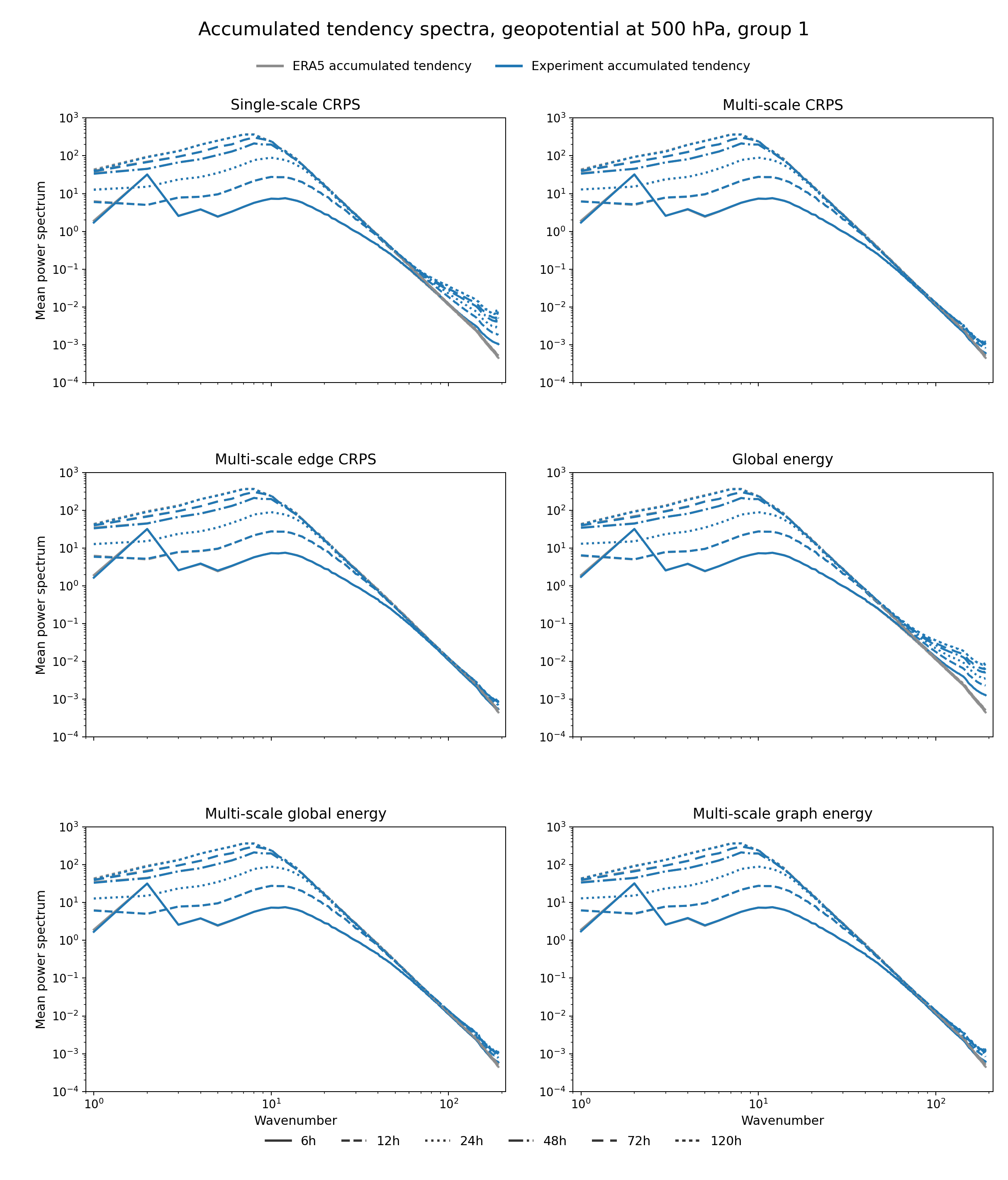}
\caption{Accumulated tendency spectra for geopotential at 500~hPa.}
\label{fig:small-accumulated-tendency-z500-spectra-g1}
\end{figure}

\begin{figure}[p]
\centering
\includeSpectraFigure{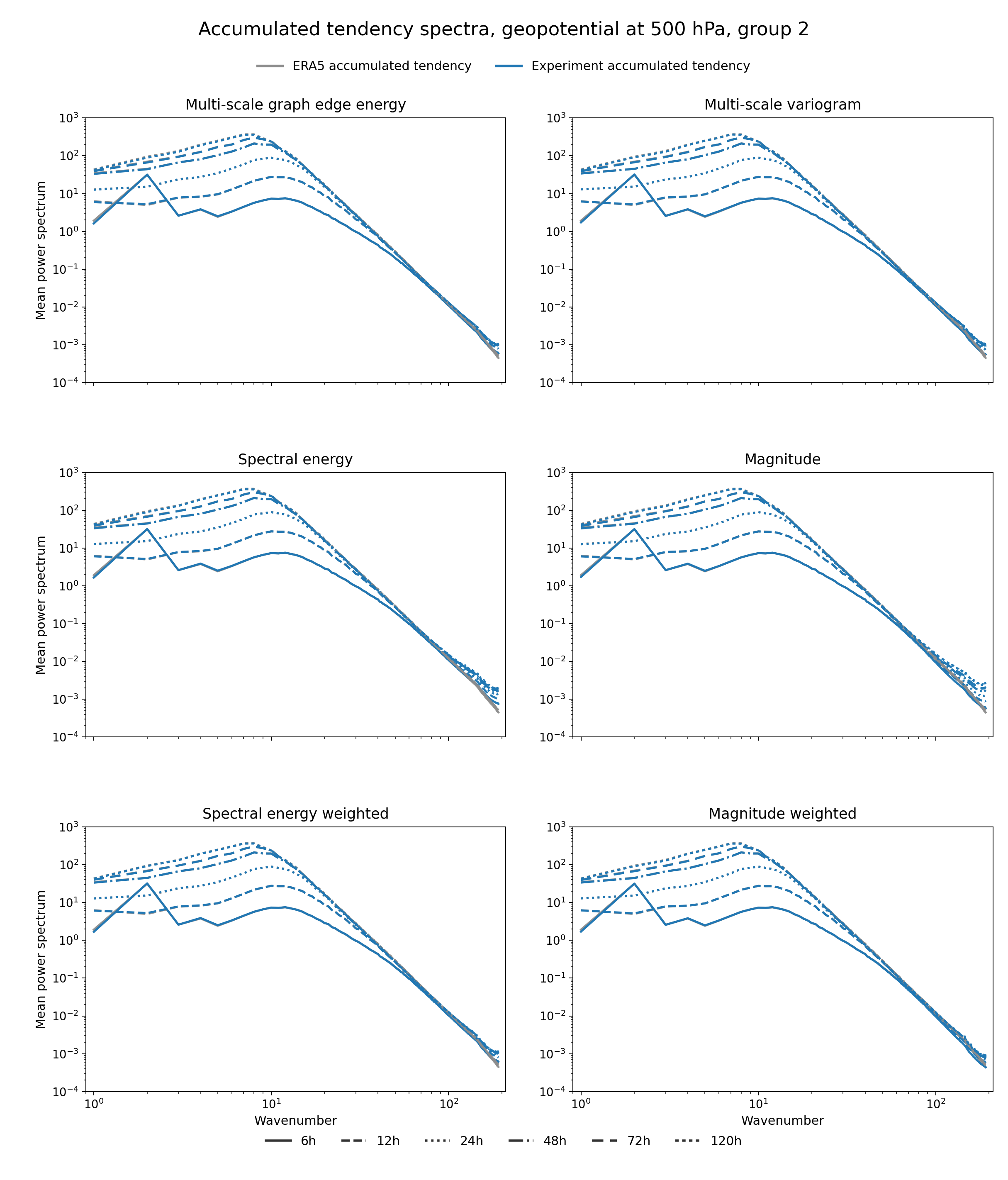}
\caption{Accumulated tendency spectra for geopotential at 500~hPa.}
\label{fig:small-accumulated-tendency-z500-spectra-g2}
\end{figure}

\begin{figure}[p]
\centering
\includeSpectraFigure{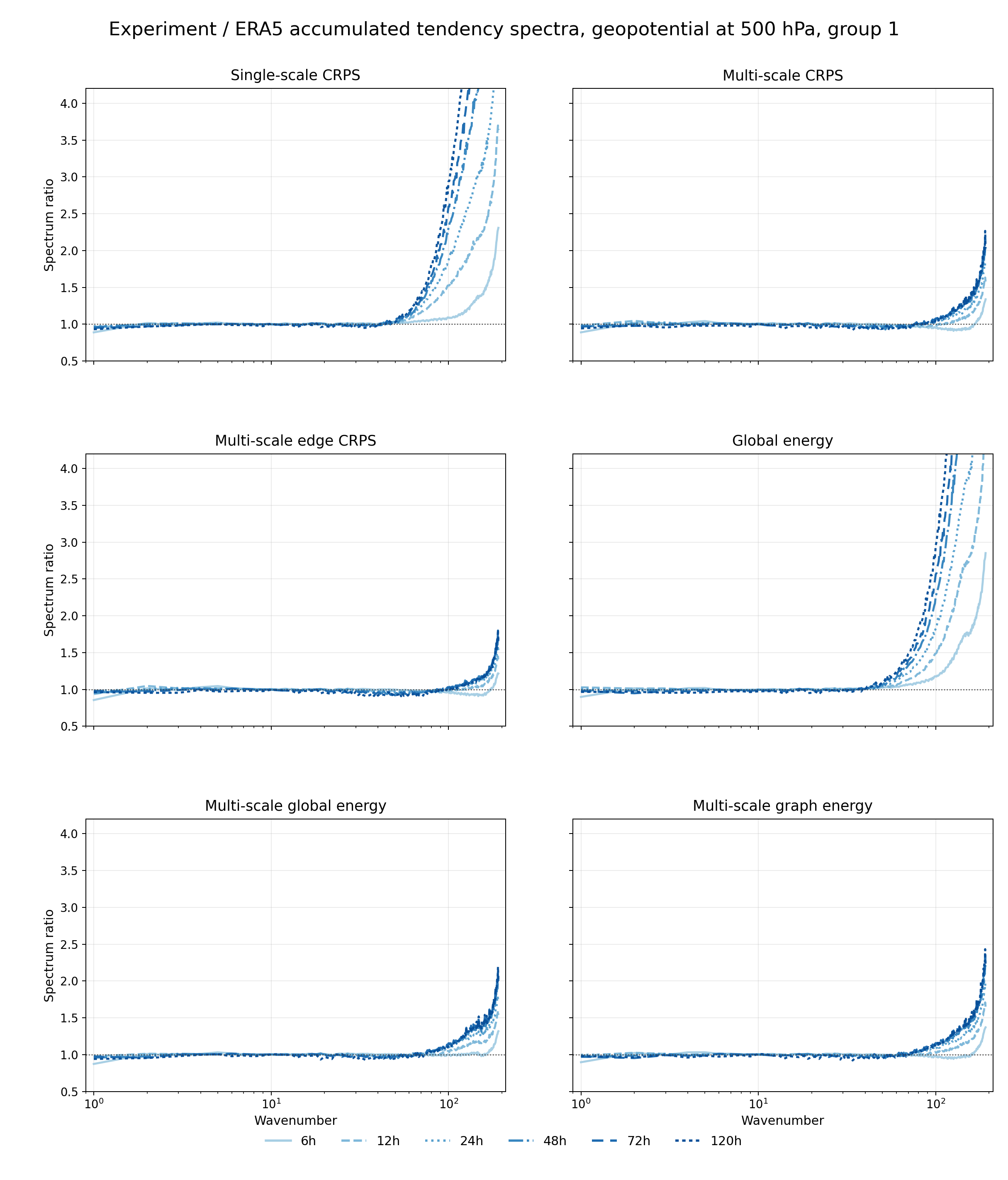}
\caption{Accumulated tendency-spectrum ratios for geopotential at 500~hPa.}
\label{fig:small-accumulated-tendency-z500-ratio-g1}
\end{figure}

\begin{figure}[p]
\centering
\includeSpectraFigure{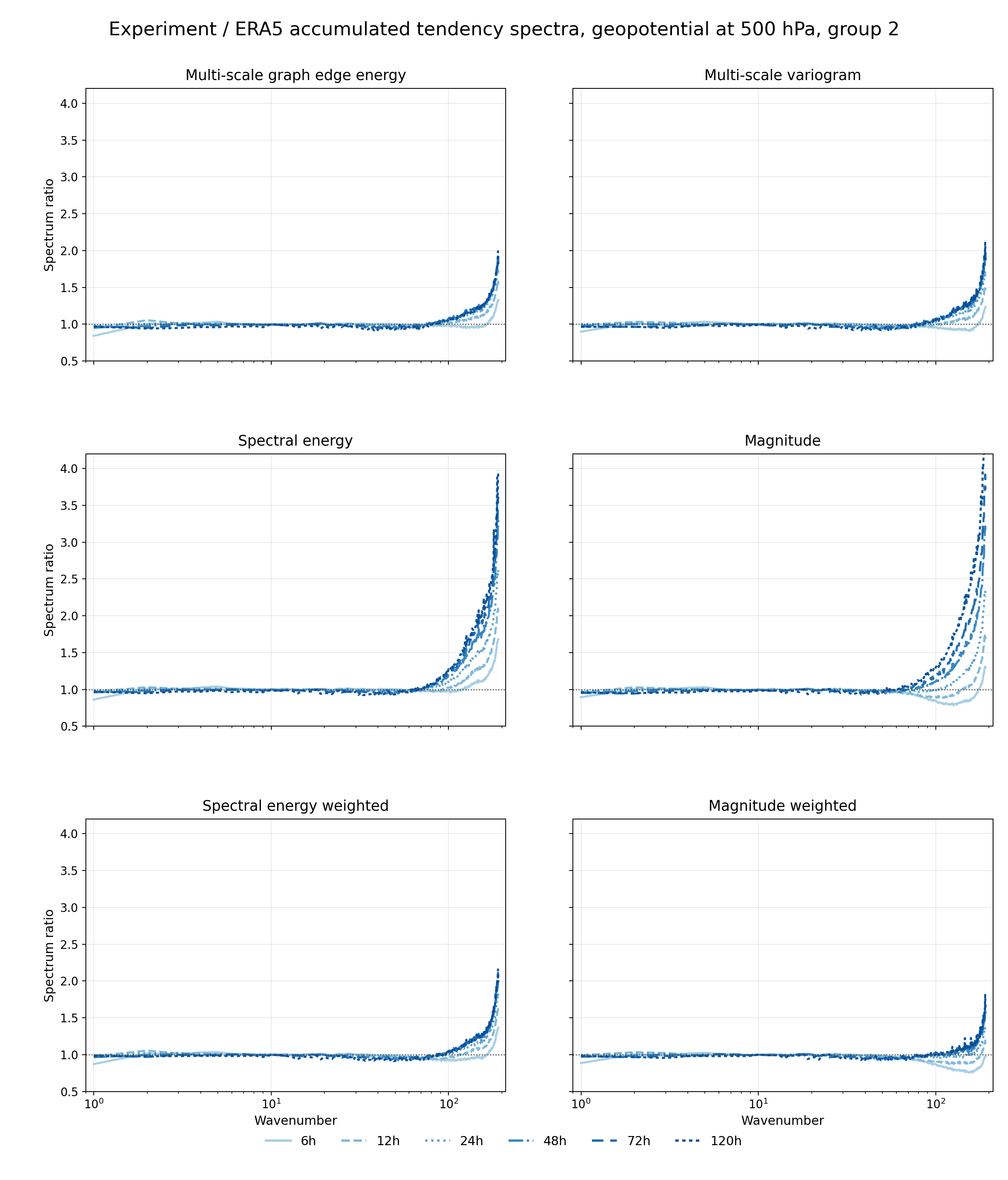}
\caption{Accumulated tendency-spectrum ratios for geopotential at 500~hPa.}
\label{fig:small-accumulated-tendency-z500-ratio-g2}
\end{figure}

\begin{figure}[p]
\centering
\includeSpectraFigure{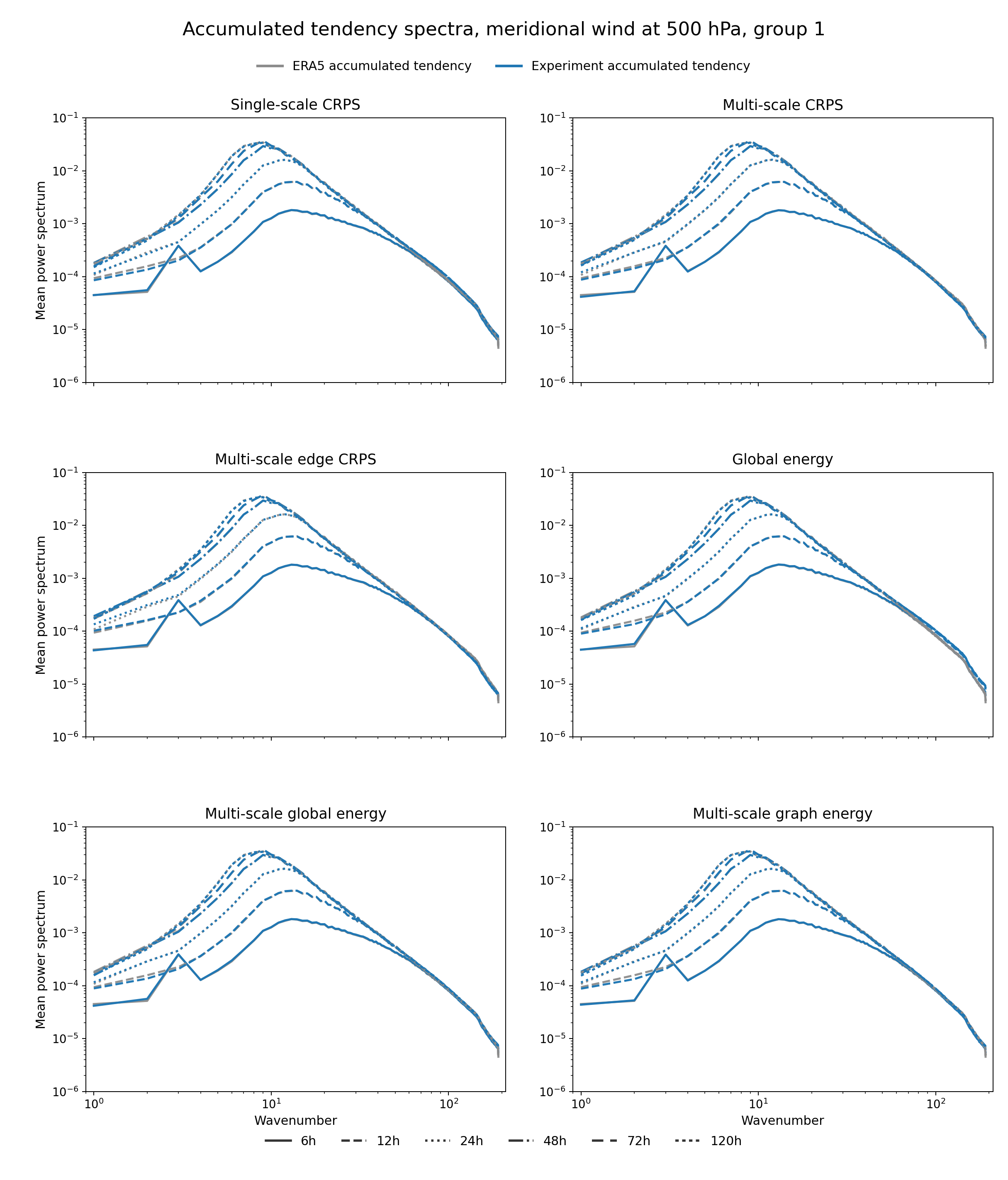}
\caption{Accumulated tendency spectra for meridional wind at 500~hPa.}
\label{fig:small-accumulated-tendency-v500-spectra-g1}
\end{figure}

\begin{figure}[p]
\centering
\includeSpectraFigure{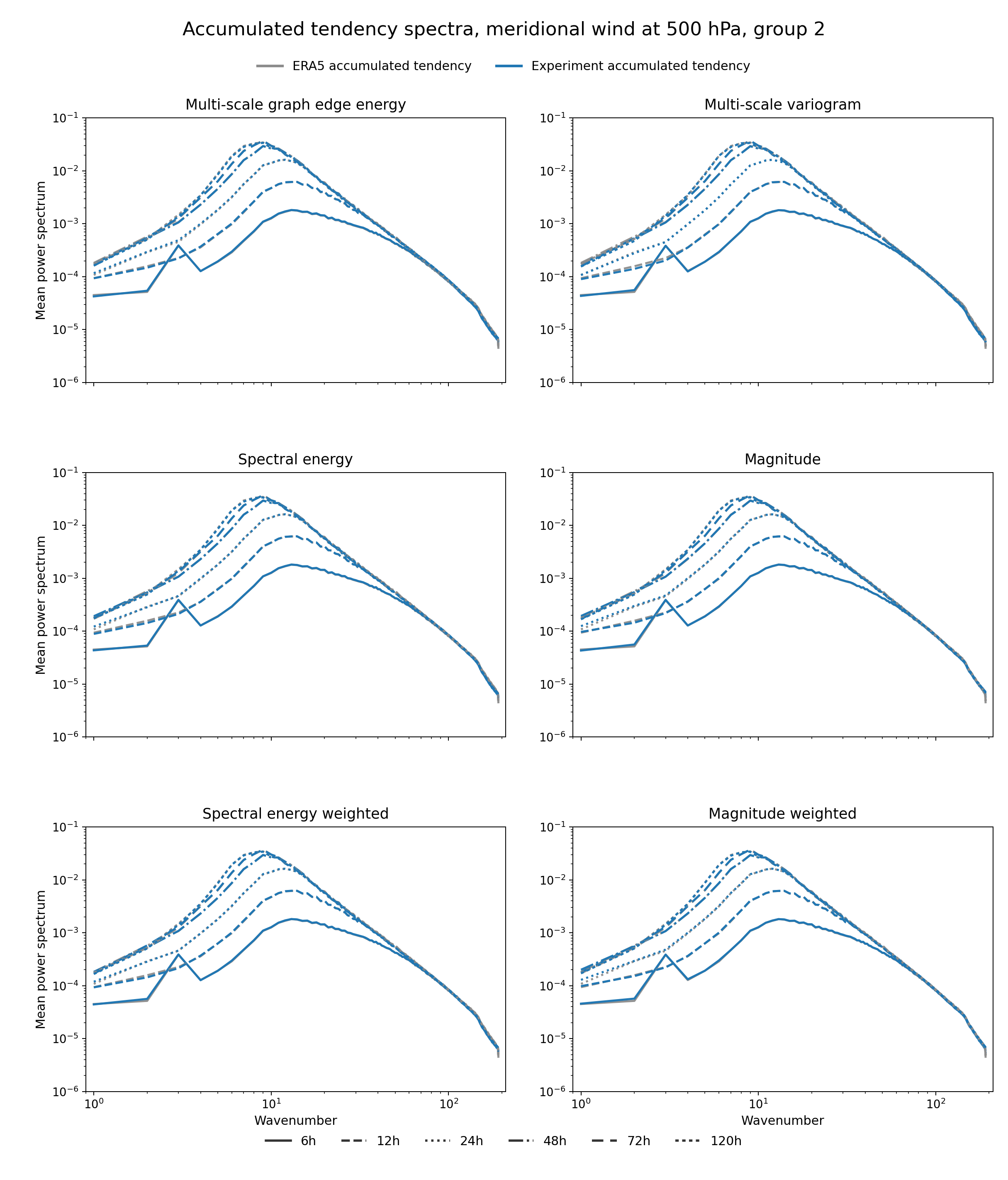}
\caption{Accumulated tendency spectra for meridional wind at 500~hPa.}
\label{fig:small-accumulated-tendency-v500-spectra-g2}
\end{figure}

\begin{figure}[p]
\centering
\includeSpectraFigure{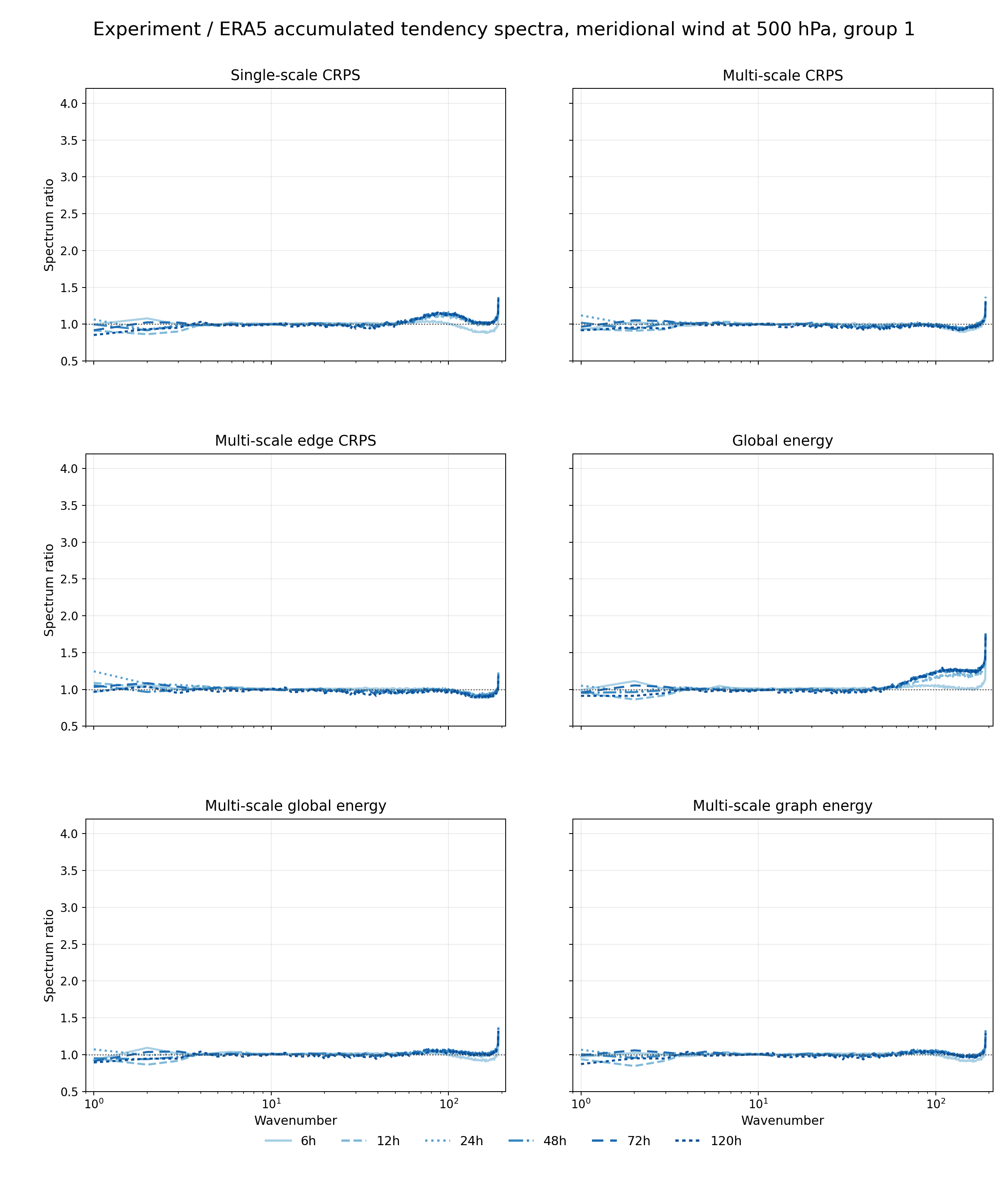}
\caption{Accumulated tendency-spectrum ratios for meridional wind at 500~hPa.}
\label{fig:small-accumulated-tendency-v500-ratio-g1}
\end{figure}

\begin{figure}[p]
\centering
\includeSpectraFigure{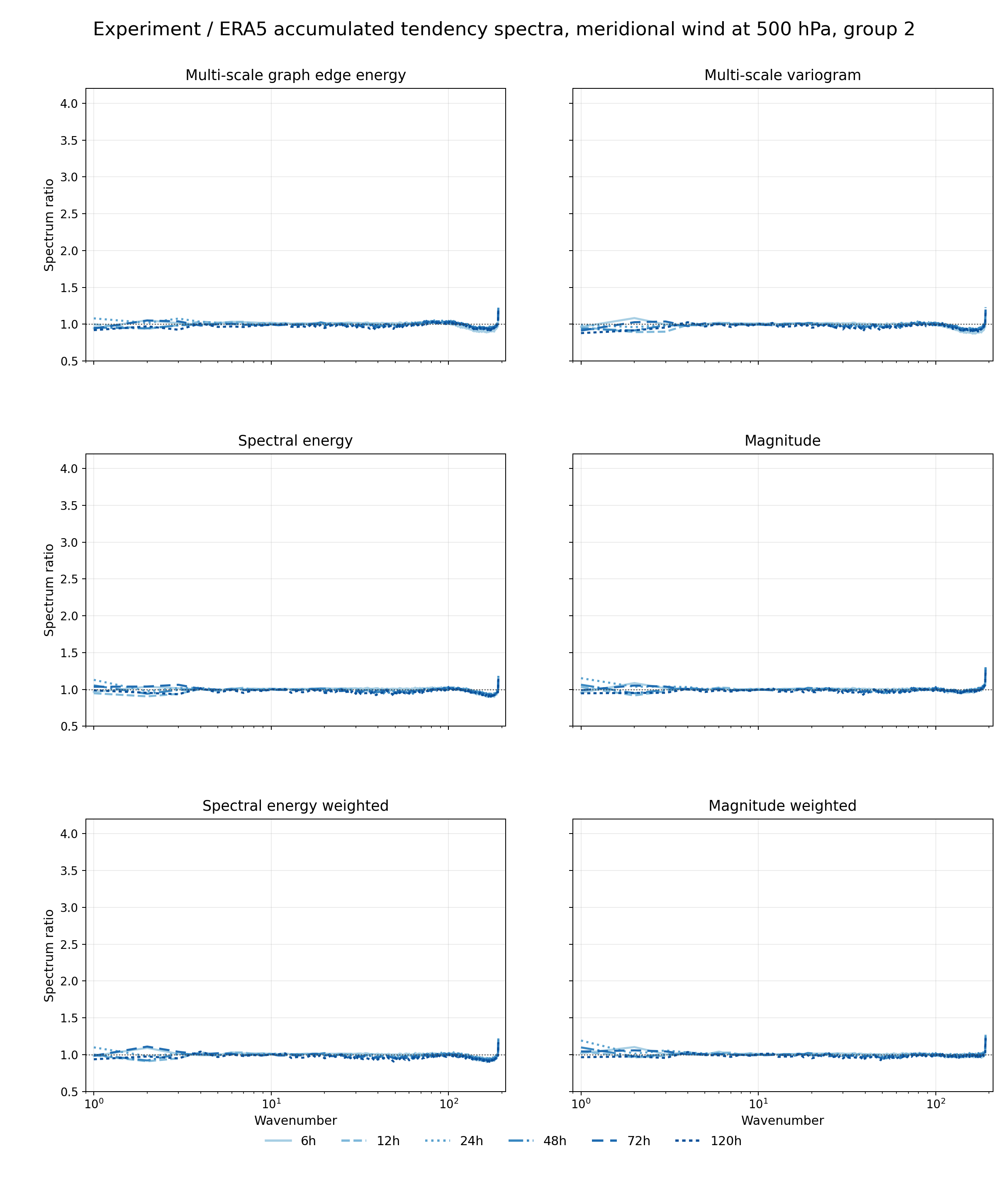}
\caption{Accumulated tendency-spectrum ratios for meridional wind at 500~hPa.}
\label{fig:small-accumulated-tendency-v500-ratio-g2}
\end{figure}

\begin{figure}[p]
\centering
\includeSpectraFigure{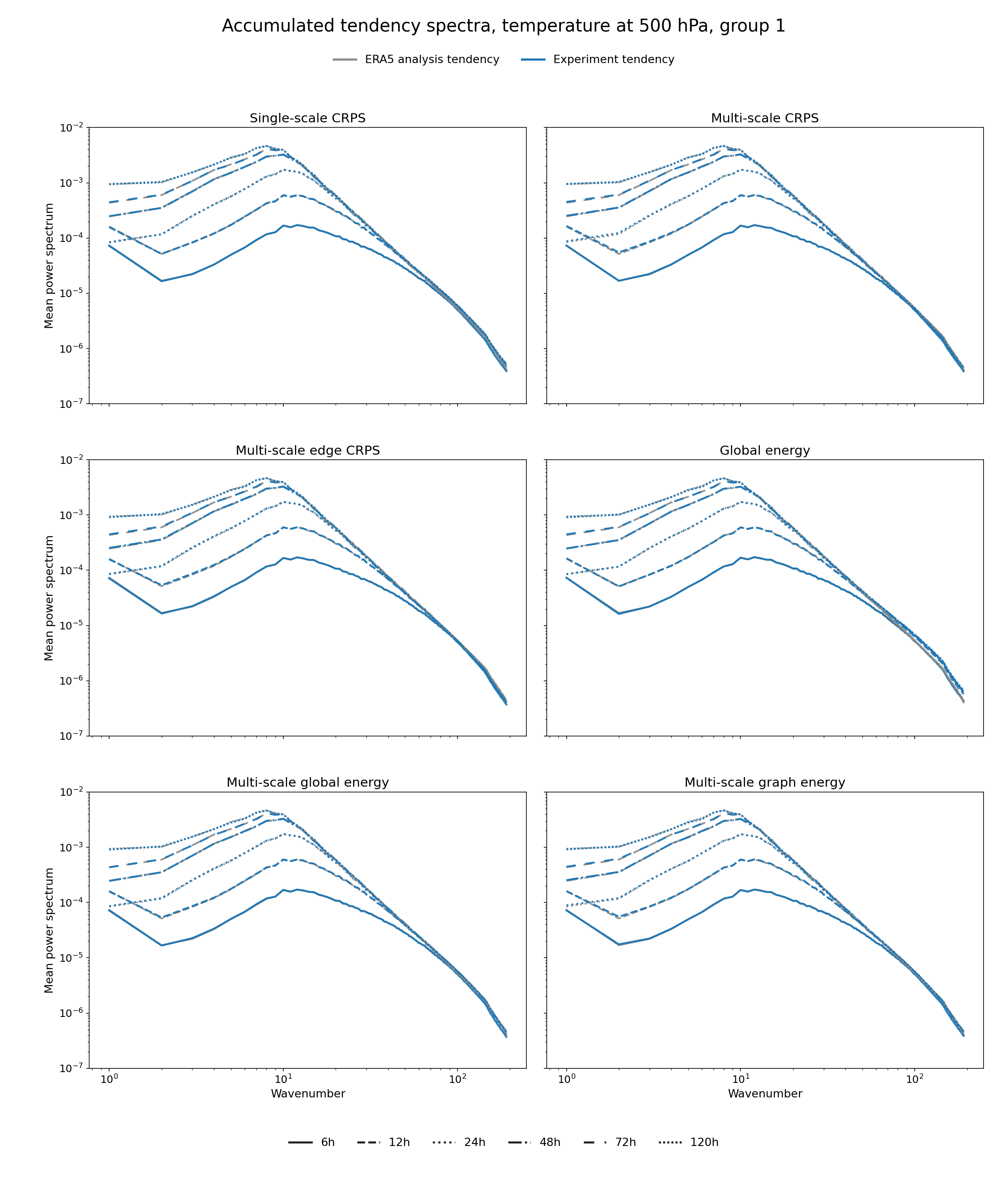}
\caption{Accumulated tendency spectra for temperature at 500~hPa.}
\label{fig:small-accumulated-tendency-t500-spectra-g1}
\end{figure}

\begin{figure}[p]
\centering
\includeSpectraFigure{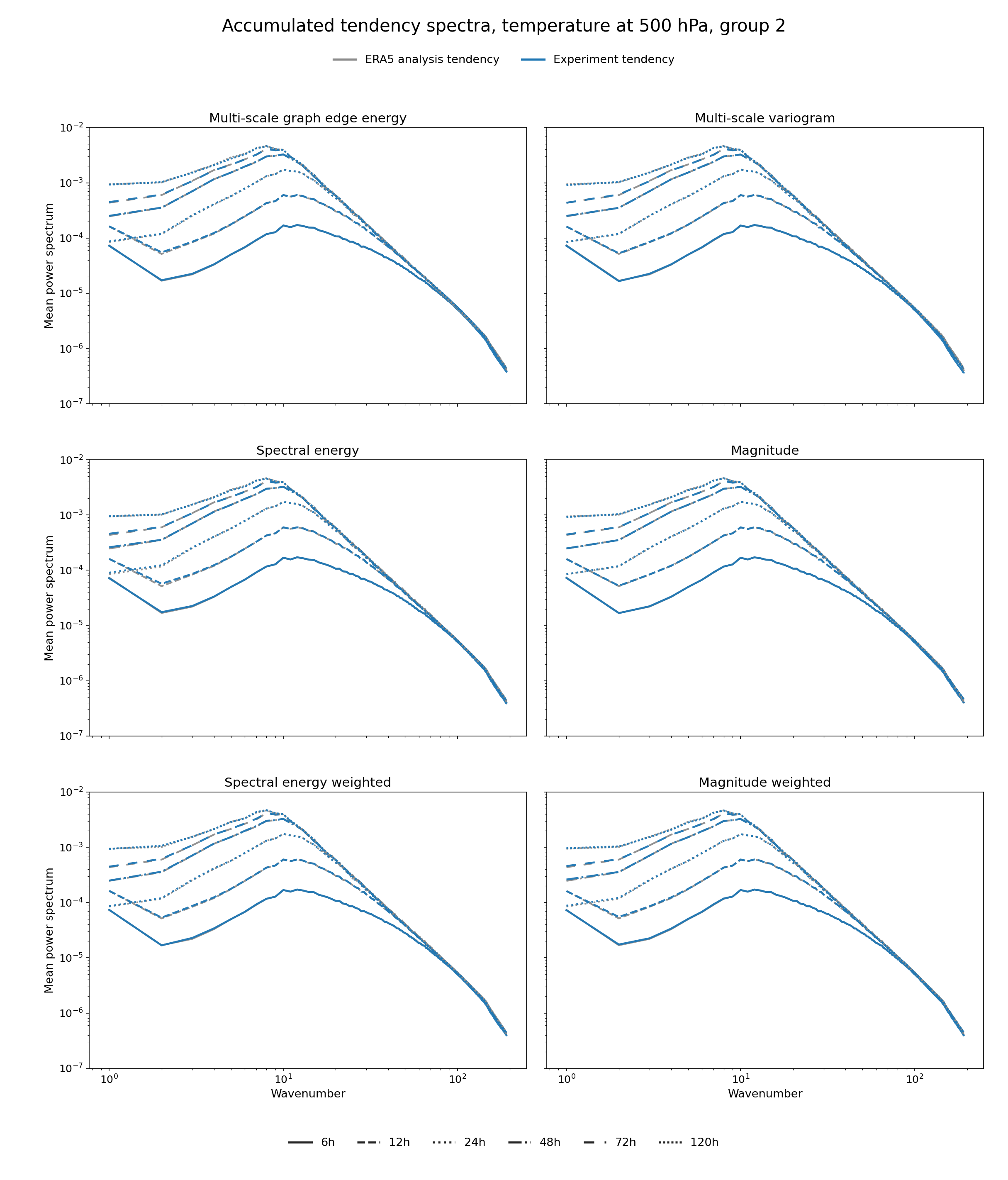}
\caption{Accumulated tendency spectra for temperature at 500~hPa.}
\label{fig:small-accumulated-tendency-t500-spectra-g2}
\end{figure}

\begin{figure}[p]
\centering
\includeSpectraFigure{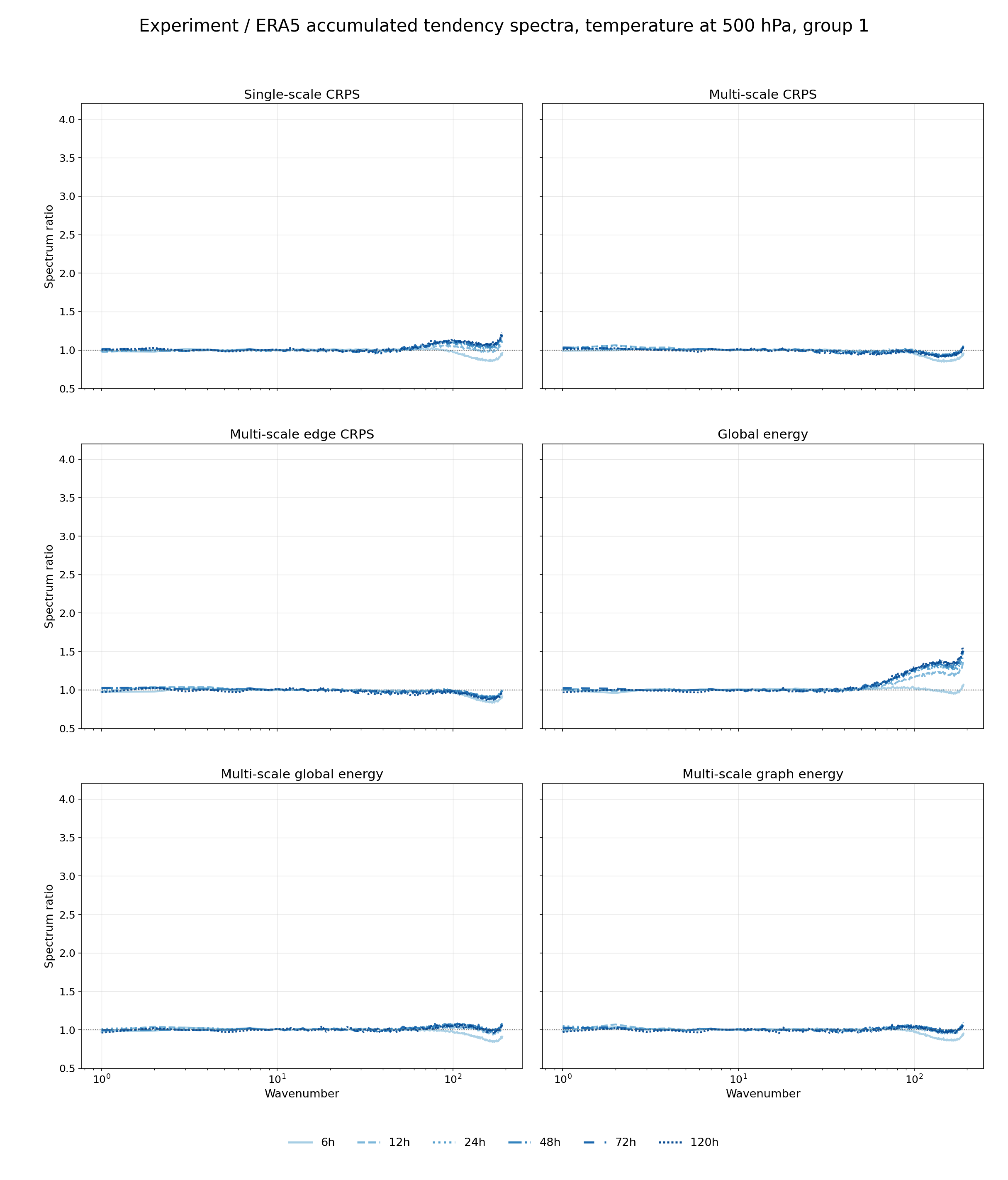}
\caption{Accumulated tendency-spectrum ratios for temperature at 500~hPa.}
\label{fig:small-accumulated-tendency-t500-ratio-g1}
\end{figure}

\begin{figure}[p]
\centering
\includeSpectraFigure{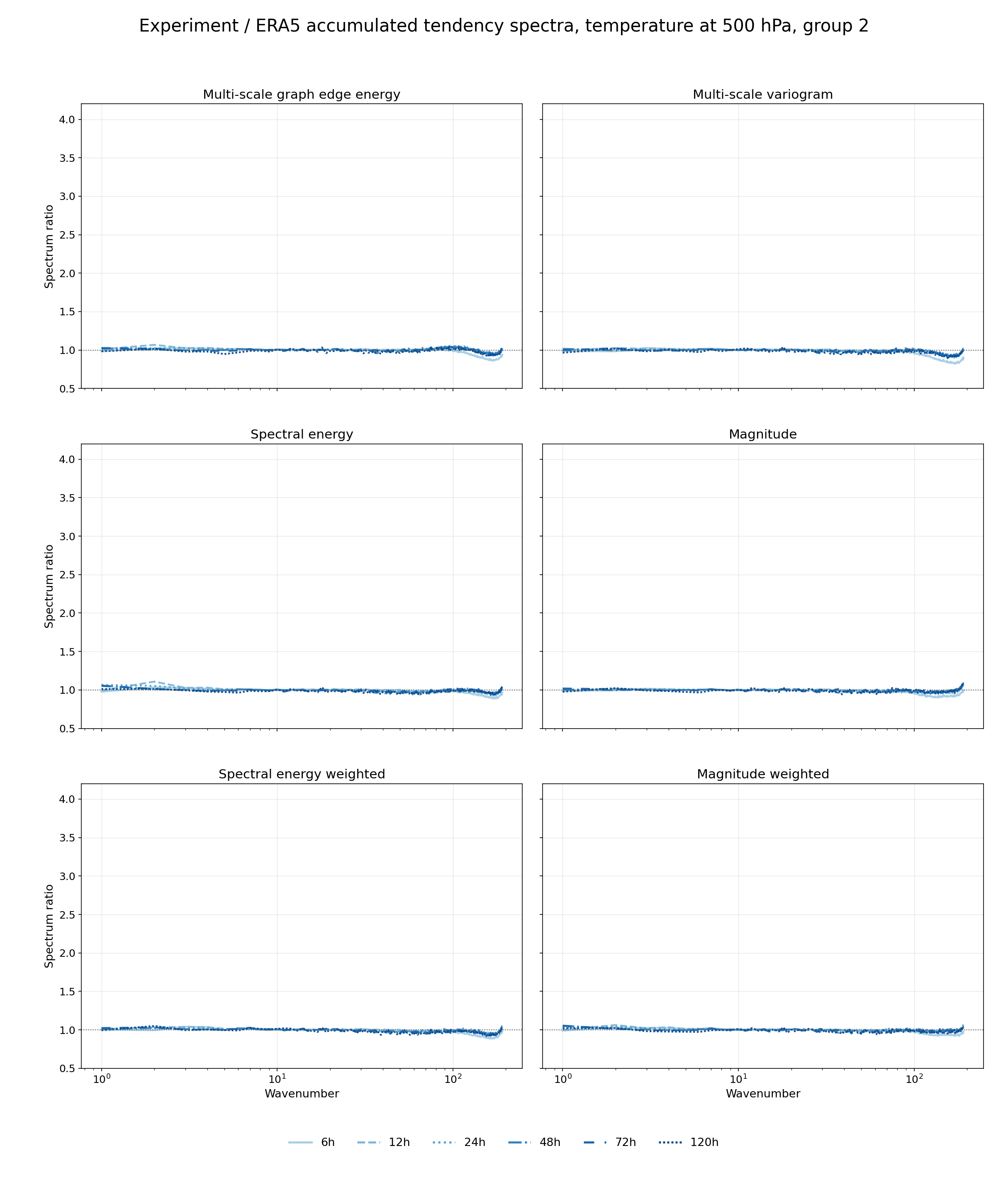}
\caption{Accumulated tendency-spectrum ratios for temperature at 500~hPa.}
\label{fig:small-accumulated-tendency-t500-ratio-g2}
\end{figure}

Different loss objectives can constrain different aspects of forecast fields. To assess the impact of the loss objectives on forecast field variability and physical realism, we compute spectra of accumulated forecast tendencies for the experiments described in section~\ref{exps2}. The accumulated tendencies are the differences between the state at a given lead time and the initial state. For ERA5, they are the differences between analyses at the corresponding valid time and initial time. Spectra are computed from 2022 forecasts initialized every four days from 1~January to 1~December at 00~UTC, 84 initialization dates in total. In addition, we compute spectra for ERA5 analyses as a reference. For the forecast fields, we average spectra over all ensemble members and use a spectral truncation of T191.  In the following, we show accumulated tendency spectra and ratios of forecasts versus ERA5, for geopotential, meridional wind, and temperature at 500~hPa (figures~\ref{fig:small-accumulated-tendency-z500-spectra-g1}--\ref{fig:small-accumulated-tendency-t500-ratio-g2}).

In these experiments, the single-scale CRPS and global energy score poorly constrain small-scale variability. The global energy score does not appear to offer an advantage over CRPS. This is consistent with previous work showing that the energy score can be relatively insensitive to correlation structure, especially in high-dimensional settings \cite{scheuerer2015variogram,pacchiardi2024probabilistic}.

Making the forecast scores scale-aware substantially improves small-scale variability and leads to more realistic forecast fields. This is true both for the multi-scale loss formalism and for spectral scores. However, scale awareness alone does not guarantee realistic variability at all scales and for all variables, as shown by the differences between spectral loss experiments with and without scale- and variable-dependent weighting: simply adding a spectral loss term constrains smaller scale variability, but not to the same degree as the multi-scale experiments that include different weighting per scale (see figure~\ref{fig:small-accumulated-tendency-z500-spectra-g1} and figure~\ref{fig:small-accumulated-tendency-z500-ratio-g1}). However, after introducing waveband weighting factors, the spectra look very similar.

In general, achieving realistic spectra for geopotential appears more challenging than realistic spectra for wind and temperature in these experiments. For all variables, differences between scale-aware loss objectives are comparatively small. The edge-CRPS experiment and the spectral magnitude CRPS experiment seem to be slightly more successful in constraining small scales for most variables than the other experiments. Some experiments show a slight overcompensation for some variables, where the early lead-time tendencies contain less small-scale variability than the ERA5 reference tendency.

\clearpage

\section{Discussion and Conclusion}
In our forecast skill experiments, all loss objectives lead to similar large-scale skill, although the tropical results show some separation, with the graph energy score experiment performing best and the energy score experiment showing some degradation. These results suggest that localized multivariate training objectives are a promising alternative to CRPS-based training, whereas the purely global energy score appears less robust in the present setup. The graph-based approach offers a flexible alternative to patch-based localization because it is not tied to fixed rectangular windows on a structured grid and can therefore be applied to irregular grids or sparse observation networks. The same approach could also be extended to local space-time neighbourhoods.

Realistic spectra of forecast fields are one measure of physical realism. Different ways have been proposed to improve spectral representation in machine-learned weather forecast models. For example, \cite{lang2024aifscrpsensembleforecastingusing} injects noise via conditional layer norms and then applies reference field truncation, where the model forecasts a tendency relative to a truncated, low resolution version of the input state. The now operational version of the AIFS ensemble uses the scale aware loss formulation introduced in \cite{lang2025multiscalelossformulationlearning}. Like AIFS-CRPS, FGN \cite{alet2025skillful} injects noise via conditional layer norms, but uses a single noise vector at all locations instead of spatially varying random fields, to encourage coherence in forecast fields. FourCastNet~3 has a spectral loss term like the spectral magnitude CRPS loss. In addition, it makes use of spatially correlated random fields, and instead of using a truncated reference state, uses its lower resolution processor grid representation interpolated to the output resolution to decode the output fields \cite{bonev2025fourcastnet3}. Spectral fidelity has also been targeted directly through attention at native resolution \cite{zhdanov2026sparseattentiondetailspreserving}.

Our results show that scale-aware losses improve spectral fidelity. The weighting of different fields, scales, and variables can matter as much as, or more than, the specific mechanism that is used to inject spatial awareness into the loss. By comparison, the choice of multivariate score has a smaller effect.

One limitation of the study is that experiments were conducted with smaller models, relatively low spatial resolution, and a shortened training schedule. Greater model capacity and longer training may improve the performance of losses without scale awareness or reduce the differences between experiments using different weightings. More testing would be needed to understand what role model architecture plays. In addition, experiments at higher resolution may yield different conclusions, as maintaining spatial coherence across forecast fields is particularly challenging at high resolution \cite{nordhagen2025highresolution}. 

Overall, multivariate scoring rules are a promising direction for generative modelling, but further research is needed to better understand how results depend on localization, scale awareness and model architecture.

\section*{Acknowledgements}
\begin{sloppypar}
The authors gratefully acknowledge the Gauss Centre for Supercomputing e.V. (www.gauss-centre.eu) for funding this project by providing computing time on the GCS Supercomputer JUPITER at Jülich Supercomputing Centre (JSC). We acknowledge the EuroHPC Joint Undertaking for awarding this project access to the EuroHPC supercomputer Jupiter, hosted by JSC in Juelich, Germany through a EuroHPC JU Special Access call. The authors also gratefully acknowledge all contributors to the Anemoi framework; in particular, we thank Sergio Portilla, Sara Hahner, and Helen Theissen for their contributions related to the spectral transform implementation in Anemoi.
\end{sloppypar}

\bibliographystyle{plain}
\bibliography{references}

\end{document}